# Mercury-related health benefits from retrofitting coal-fired power plants in China

Jiashuo Li[a], Sili Zhou[b], Wendong Wei[c], Jianchuan Qi[d], Yumeng Li[d], Bin Chen[e], Ning Zhang[a], Dabo Guan[f], Haoqi Qian[g], Xiaohui Wu[d], Jiawen Miao[h], Long Chen[I], Sai Liang[d], Kuishuang Feng[a]

[a]Institute of Blue and Green Development, Shandong University, Weihai, 264209, P.R. China
[b]State Key Laboratory of Coal Combustion, School of Energy and Power Engineering, Huazhong University of Science and Technology, Wuhan, 430074, P.R. China
[c]School of International and Public Affairs, Shanghai Jiao Tong University, Shanghai, 200030, P. R. China
[d]State Key Joint Laboratory of Environment Simulation and Pollution Control, School of Environment, Beijing Normal University, Beijing, 100875, P. R. China
[e]Laboratory of Anthropogenic Systems Ecology, College of Engineering, Peking University, Beijing, 100871, P. R. China
[f]Department of Earth System Science, Tsinghua University, Beijing, 100871, P. R. China
[g]Institute for Global Public Policy, Fudan University, Shanghai, 200433, P. R. China
[h]School of Optical-Electrical and Computer Engineering, University of Shanghai for Science and Technology, Shanghai, 200093, P. R. China
[I]Key Laboratory of Geographic Information Science (Ministry of Education), School of Geographic Sciences, East China Normal University, Shanghai, 200241, P. R. China

## ABSTRACT

China has implemented retrofitting measures in coal-fired power plants (CFPPs) to reduce air pollution through small unit shutdown (SUS), the installation of air pollution control devices (APCDs) and power generation efficiency (PGE) improvement. The reductions in highly toxic Hg emissions and their related health impacts by these measures have not been well studied. To refine mitigation options, we evaluated the health benefits of reduced Hg emissions via retrofitting measures during China's 12[th] Five-Year Plan by combining plant-level Hg emission inventories with the China Hg Risk Source-Tracking Model. We found that the measures reduced Hg emissions by 23.5 tons (approximately 1/5 of that from CFPPs in 2010), preventing 0.0021 points of per-foetus intelligence quotient (IQ) decrements and 114 deaths from fatal heart attacks. These benefits were dominated by CFPP shutdowns and APCD installations. Provincial health benefits were largely attributable to Hg reductions in other regions. We also demonstrated the necessity of considering human health impacts, rather than just Hg emission reductions, in selecting Hg control devices. This study also suggests that Hg control strategies should consider various factors, such as CFPP locations, population densities and trade-offs between reductions of total Hg (THg) and $Hg^{2+}$.

# INTRODUCTION

Hg is a globally recognized contaminant posing great risks to both humankind and ecosystems [1-3]. Once emitted, Hg is easily converted into methylmercury (MeHg), a highly toxic substance that causes untreatable infant IQ decrements and adult cardiovascular disease via bioaccumulation[4-6]. Scientific evidence shows that Hg pollution is the culprit of Minamata Disease, one of the major environmental disasters in the 20$^{th}$ century. Unfortunately, human health is still seriously threatened by the adverse effects of Hg. In 2010, 7360 deaths from fatal heart attacks were related to the intake of MeHg in China[3]. To protect our society from Hg-related hazards, 128 countries, including China, signed a legally binding international treaty, the *Minamata Convention on Mercury*, aiming at Hg control, which officially came into effect in 2017.

Coal-fired power plants (CFPPs) are one of the largest anthropogenic sources of global atmospheric Hg emissions. According to the Global Mercury Assessment 2013 and 2018, CFPPs contributed 16% and 14% of the global anthropogenic Hg emissions in 2010 and 2015, respectively, while in some areas, such as east and southeast Asia, Hg emissions from CFPPs witnessed a growth of 19.2% during the same period[2]. Thus, the *Minamata Convention on Mercury* lists Hg emissions from CFPPs as one of its key mitigation targets. China, as the world's largest coal-fired power generator, contributed for approximately one-third of airborne Hg from global CFPPs[7,8]. Moreover, a large number of CFPPs in China are located in densely populated areas, leading to an increase in Hg exposure risks. In particular, CFPPs are responsible for the elevation of MeHg exposure in many areas in China[9]. Given that, reducing Hg emissions from CFPPs has been put into China's political agenda.

In addition to Hg, CFPPs are major sources of emissions of $CO_2$ and pollutants such as $PM_{2.5}$, $SO_2$ and $NO_x$. The control of $CO_2$ and air pollution from CFPPs has been prioritized in China's national political agendas because China has been confronted with mounting international and domestic pressures to mitigate $CO_2$ emissions and improve ambient air quality[10]. China's central and local governments have been retrofitting CFPPs in the context of the energy revolution. These retrofitting measures can be generally categorized into three types: small unit shutdown (SUS, aimed at units with a capacity less than 300 MW)[11], the installation of efficient APCDs, and PGE improvement (via upgrading generation technologies). Studies have found that the implementation of these retrofitting measures can reduce Hg emissions simultaneously[12,13]. According to the articles of the *Minamata Convention*, all the parties are required to regularly facilitate the evaluation of mitigation measure effectiveness and provide an assessment of the related impacts on human health. Thus, evaluating the co-benefits of reducing Hg emissions and associated health impacts from CFPP retrofitting is vital for both fulfilling the international obligation and formulating refined mitigation policies.

Scholars have made initial attempts to investigate the impacts of CFPP retrofitting on Hg emissions. For instance, a recent study found that improving APCD's mercury removal efficiency substantially reduced THg emissions (over 40 t) from China's coal power sector during 2013-2017[14,15]. However, these studies take all the CFPPs as a whole and cannot explicitly reflect the significant heterogeneity in Hg emissions from individual CFPPs (caused by their strikingly different Hg content in coal and APCD types). Although Liu et al.[16] compiled a high-resolution inventory on Hg from 1817 CFPPs in China based on the detailed APCD parameters, the plant-level retrofitting measures' effect on Hg emission changes remains poorly understood. Moreover, existing studies on Hg-related health impacts are at the national or provincial scale[1,3]. As the real world retrofitting actions occurred at the plant-level, a precise assessment comprehensively considering factors such as local population density and Hg deposition is thus in urgent need to reflect each CFPP's contribution to health benefits.

Here, we combined plant-level Hg emission inventories and the China Hg Risk Source-Tracking Model (CMSTM)16 to quantify the Hg-related health impacts from China's CFPP retrofitting campaign during the 12th Five Year Plan (FYP) period (details in Methods). We for the first time present a high-resolution map of both Hg reduction and related health benefits associated with the CFPP retrofitting measures. We compare the benefits of various retrofitting measures, and make targeted suggestion to CFPP retrofitting in different areas by comprehensively considering factors such as CFPP locations and population densities. The findings of this study provide not only a scientific foundation for monitoring the implementation of the Minamata Convention on Mercury in China's coal fired power sector but also useful information for more refined mitigation strategies.

**Emission reductions from CFPP retrofitting measures.** The three CFPP retrofitting measures resulted in an overall emission reduction of 23.51 t (72.38% $Hg^0$, 26.82% $Hg^{2+}$ and 0.80% $Hg_p$) during 2011-2015, equivalent to approximately 20% of the THg emitted by CFPPs in 2010. Hg emission reduction occurred in 29 out of the 34 provincial regions, among which Jiangsu, Inner Mongolia and Shandong had the top three greatest reductions (Figure 1(a)). Among the three retrofitting measures, over half of the national total emission reduction was attributed to the newly installed APCDs, while SUS and PGE had lower contributions. The contributions of the three types of measures in different provinces are shown in Figure 1(a). Meanwhile, China's top five state-owned power generation groups (Huaneng Group, Datang Corporation, Huadian Corporation, Guodian Corporation and State Power Investment Corporation), accounting for 46.29% of China's total CFPP capacity, contributed more than half of the emission reductions. The rest of the reductions were made by other large state-owned enterprises, local enterprises (owned by provincial or municipal governments) and private enterprises.

Among the three measures, the increasing APCD installation rate made the largest contribution to emission reductions. As illustrated in Figure 1(b), 308 CFPPs in 24 provinces were equipped with new pollution control devices such as wet flue gas desulfurization (WFGD) and selective catalytic reduction (SCR) technology, which helped avoid 12.02 tons of atmospheric Hg emissions (81.69% $Hg^0$, 17.82% $Hg^{2+}$ and 0.49% $Hg_p$, see SI Table S1 for more details). Notably, some of the newly installed APCDs reduced the THg emissions but also led the growth of $Hg^{2+}$ or $Hg_p$ emissions. For example, as SCR was deployed in many CFPPs, $Hg^{2+}$ emissions in Anhui and Guizhou increased by 0.09 tons and 0.04 tons, respectively. This is because SCR can change the speciation profile of Hg emissions in flue gas by oxidizing $Hg^0$ to $Hg^{2+}$[16,17]. SUS contributed 9.19 tons of Hg emission reductions nationwide (SI Table S1), only second to the reductions attributed to newly installed APCDs. The power generation capacity of the decommissioned CFPPs was equivalent to 5% of the national total in 2010, while the associated reduction in Hg emissions was equivalent to 7.75% of the total emissions from China's CFPPs in 2010, indicating that the decommissioned units were more Hg intensive than the national average level. Improving PGE via the recovery and utilization of waste heat is also an effective way to reduce Hg emissions. During the 12th FYP, the national average coal consumption rate (the amount of coal consumption per kilowatt-hour of electricity generated) of China's CFPPs decreased from 312 g/KWh to 297 g/KWh[18,19]. Such PGE reduced atmospheric Hg emissions by 2.3 tons (data for each CFPP are shown in SI Table S1).

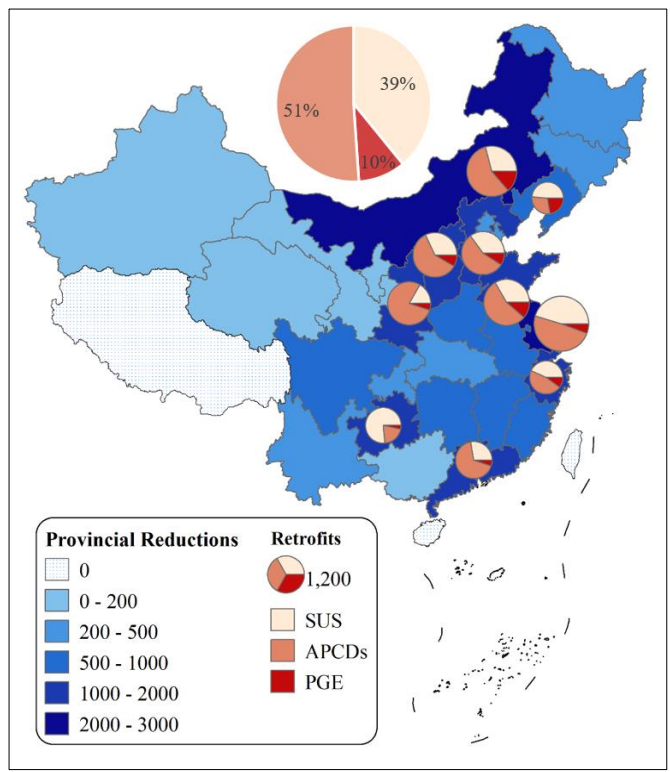 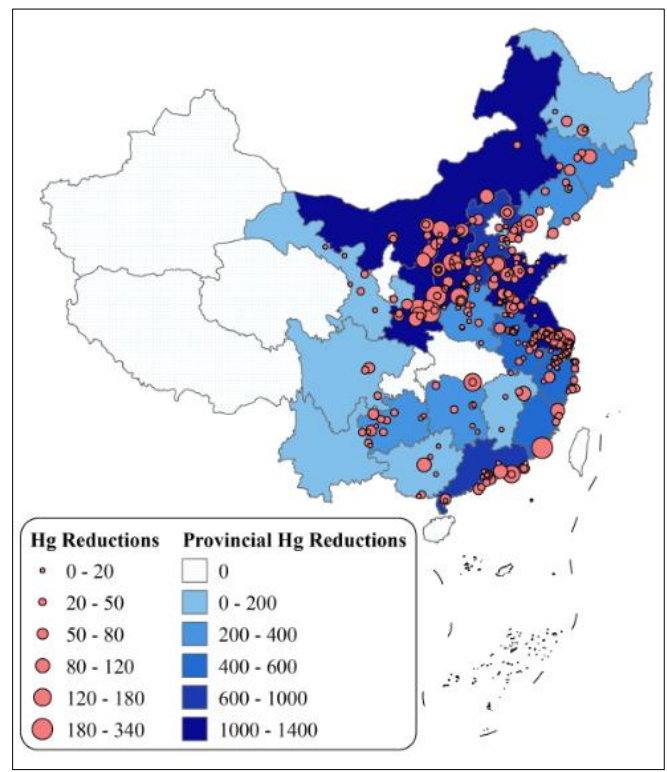

(a) Total  (b) APCDs

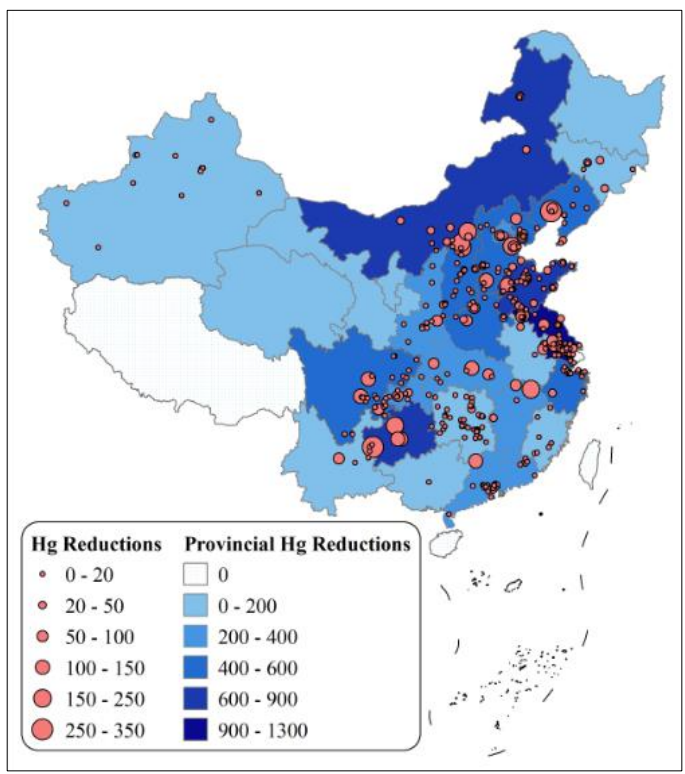 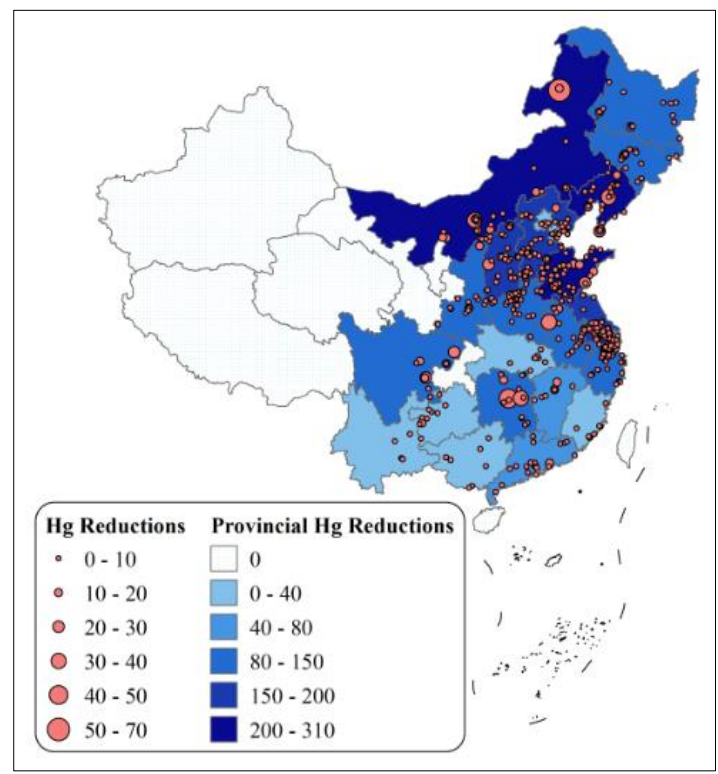

(c) Small unit shutdown  (d) Efficiency improvement

Figure1. Reductions in atmospheric Hg emissions by three types of retrofitting measures (unit: kg)

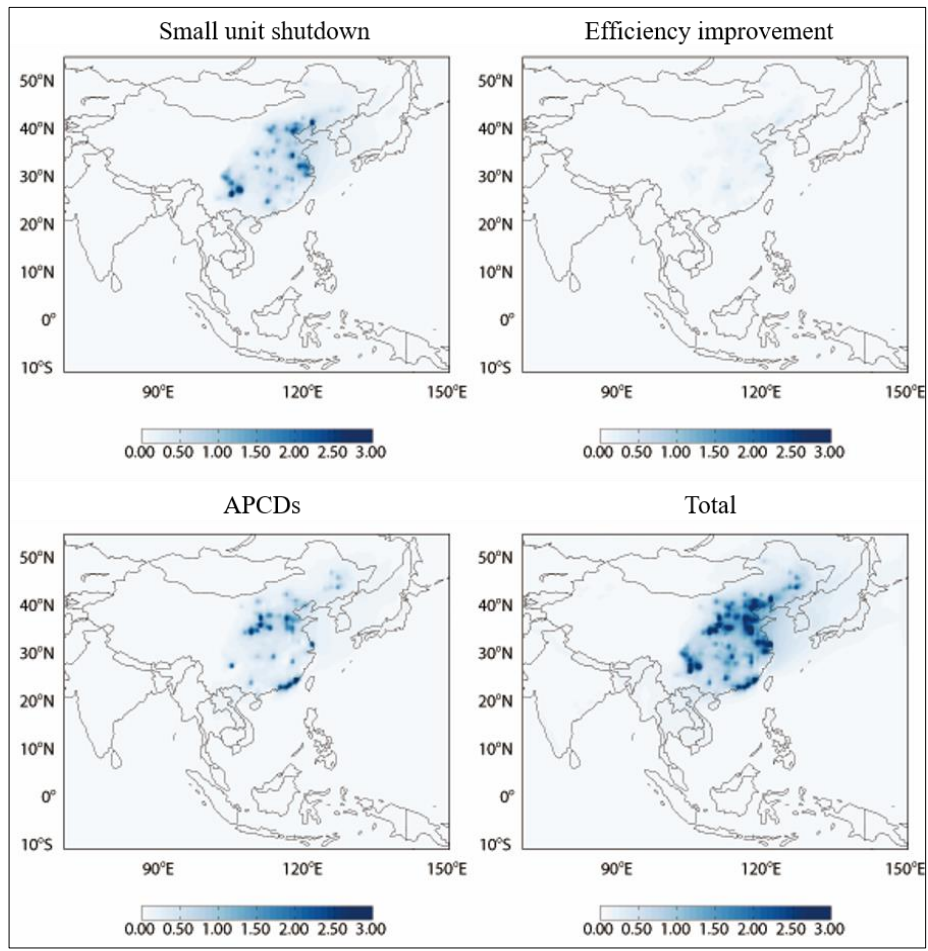

Figure 2. The Hg deposition reductions by the three retrofitting measures (unit: kg)

**Reduced atmospheric Hg deposition.** The three retrofitting measures prevented 5.10 tons of atmospheric Hg deposition in China, of which 2.38 tons for CFPP shutdown, 2.17 tons for APCD installation and 0.55 tons for PGE. Figure 2 illustrates the spatial distributions of the reductions in atmospheric Hg deposition over China. Among all provincial regions, the Hg deposition reduction was the greatest in Inner Mongolia (0.38 t), followed by Shandong (0.36 t) and Hebei (0.35 t). It is interesting to note that there is a significant mismatch between the Hg emission reduction and deposition in many regions. Although the reduction in Hg emissions is higher in Jiangsu (0.26 t) than in Inner Mongolia, the reduction in atmospheric Hg deposition over Inner Mongolia is 1.5 times greater than that over Jiangsu. The explanation for this phenomenon is that Inner Mongolia has a much larger territory than Jiangsu and therefore experienced a greater reduction in atmospheric Hg deposition from other regions. The air transport effect also played a vital role in the reduction of Hg depositions in each province. It was found that, for the provinces with the top ten largest Hg reductions, more than half of the reductions in Hg depositions resulted from decreases in the Hg transported from other provinces, especially from their neighbouring provinces (summarized in Table S2). For example, Hg deposition in Hebei decreased by 0.35 tons, but less than one-third of this decrease was derived from local CFPPs within Hebei's territory. The largest trans-regional contributors to deposition reductions in Hebei were Shandong (0.08 tons), Shanxi (0.04 tons) and Inner Mongolia (0.04 tons). Although there was no reduction in Hg emissions in Xizang, the province experienced a decline of 0.06 tons in Hg deposition due to the emission reduction measures of CFPPs in other provincial regions.

**The health benefits from CFPP retrofitting.** CFPP retrofitting prevented 0.0021 points of per-foetus intelligence quotient (IQ) decrements and 114 deaths from fatal heart attacks (points and deaths hereafter), equivalent to 9.09% and 9.26% of the total IQ decrements and deaths from fatal heart attacks caused by the production of electricity and heat power in China[3]. As illustrated in Figure 3, the health benefits in each province had large spatial variability (see details in SI Table S3), and approximately 70% of the contributions (0.0015 points and 78 deaths) came from the top ten provincial regions with the greatest emission reductions.

Among the three emission reduction measures, the largest health benefits came from SUS, which contributed to the prevention of 0.001 points of per-foetus IQ decrements and 59 deaths, respectively. It is interesting to note that the SUS-associated reduction in Hg depositions was greater in Inner Mongolia than in Jiangsu, but the number of avoided deaths associated with SUS in Jiangsu was more than tenfold greater than that in Inner Mongolia, this is because Jiangsu has a much higher population density and MeHg concentration in consumed food[3]. Although Hg emissions were more reduced by the installation of APCDs than by SUS, the former contributed less to health benefits (0.0008 points and 43 deaths) than the latter. The health benefits associated with newly installed APCDs were undermined by devices such as SCR, which remove THg emissions from flue gas and change the Hg species by increasing the proportions of $Hg^{2+}$ and $Hg_P$ that are easily deposited, thus adversely affecting human health[1]. Therefore, an undesirable situation occurred in Anhui Province: the health risks related to Hg pollution increased despite deceasing THg emissions. Compared with SUS and the installation of APCDs, PGE improvement had much less of a positive contribution to health benefits (0.0002 points and 12 deaths). Regarding the health benefits gained by Hg reduction through efficiency improvement nationwide, only Zhejiang and Sichuan avoided more than 1 death.

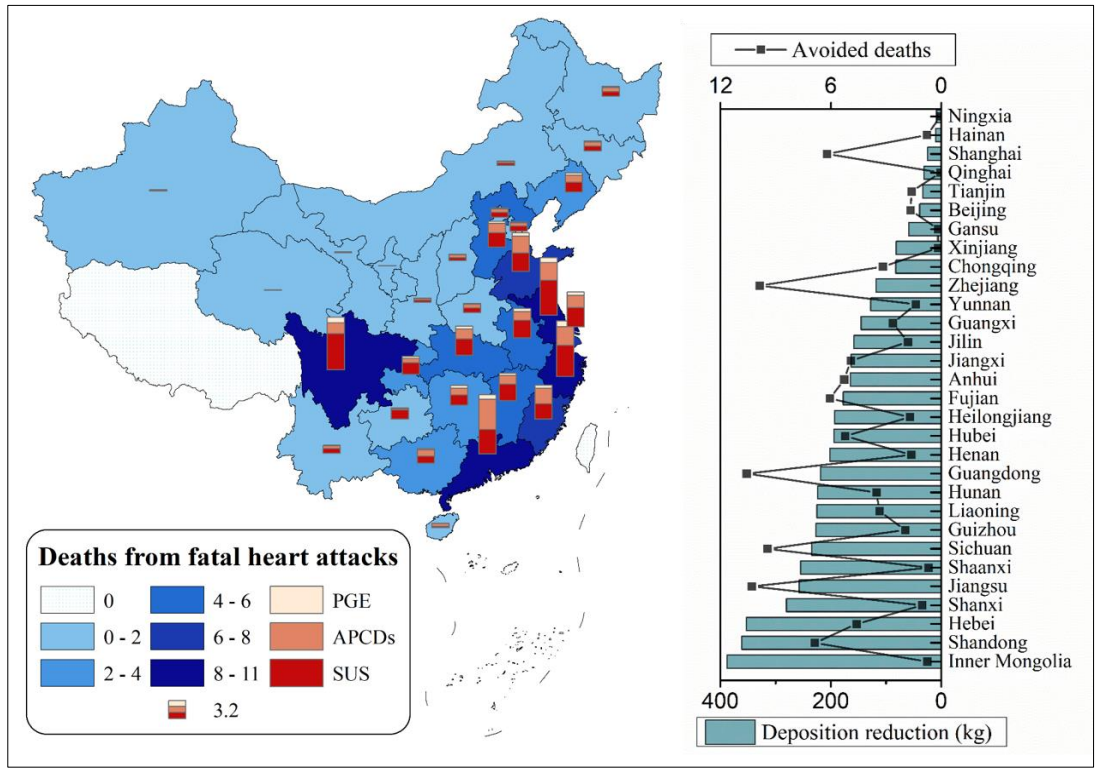

Figure 3 The spatial distribution of avoided deaths from fatal heart attacks over China and contributions of the three types of CFPP retrofitting measures

**Health benefits by power generation capacities.** Most of the reductions in health impacts associated with PGE (avoided IQ decrements and fatal heart attacks: 0.00002 points and 9 deaths) and APCD installation (0.00043 points and 39 deaths) were attributed to large CFPPs with capacities over 300 MW, accounting for 74.36% and 89.80% of the total reductions, respectively. In particular, CFPPs with capacities over 1200 MW contributed more than half of the health benefits associated with the installation of APCDs. In contrast, the CFPPs with capacities under 100 MW accounted for only approximately 8% and 3% of the health benefits related to PGE improvement and the installation of APCDs, respectively. The CFPP shutdown campaign showed a completely different picture. As this campaign mainly targeted small units, almost two-thirds of the health benefits gained by CFPP decommission (0.0002 points and 37 deaths) were from CFPPs with capacities less than 300 MW.

**Health benefits by power generation groups.** The CFPPs from the five power generation groups were the leading contributors to the observed health benefits and were responsible for half and two-fifths of the health benefits associated with the newly installed APCDs and decommissioned CFPPs, respectively. Among the five groups, the Huaneng Group made the largest contribution (avoided IQ decrements and fatal heart attacks: 0.0003 points and 18 deaths) to the health benefits, especially for the health benefits of newly installed APCDs. Furthermore, an IQ decrement of 0.0002 and 10 deaths were avoided by SUS of the Guodian Corporation, ranking first among the five groups. The local CFPPs in different regions were the second largest contributor, responsible for approximately 30% of the health benefits of SUS and 20% of the health benefits of the other two retrofitting measures. Moreover, half of the decommissioned CFPPs with capacities less than 100 MW belonged to local enterprises, which have higher emission intensity and lower energy efficiency. The rest of the health benefits were attributed to individuals and private enterprises, captive power plants and other large state-owned enterprises (i.e., China Resources Group and State Grid Corporation of China). Notably, not all CFPP retrofitting measures resulted in health benefits; for instance, health impacts slightly increased because of the newly installed SCR device of a captive CFPP-Taigang stainless steel power plant.

**Trans-regional effects for health benefits.** As the sources of health benefits related to reductions in Hg deposition were clearly identified in the atmospheric transport model, the transregional effects on health benefits (namely, the source-receptor relationship of health benefits between provincial regions) were illustrated (Figure 4). The three provinces that received the largest cross-boundary health benefits were Zhejiang, Guangdong and Jiangsu, while Shandong, Jiangsu and Hebei served as the largest contributors to other provinces. For example, 5 out of the 10 deaths avoided in Jiangsu Province were attributable to CFPP retrofitting measures in other provincial regions, such as Shandong (14.44%), Hebei (5.74%) and Inner Mongolia (4.53%). The majority of deaths avoided in Shandong (55.05%) and Hebei (71.69%) were credited to the implementation of the three types of CFPP retrofitting efforts in other provinces. The retrofitting measures in Shandong prevented 13 deaths in other provinces, accounting for over 80% of the total contributions of Shandong, and the largest beneficiaries were Jiangsu (9.19%), Zhejiang (8.91%) and Shanghai (8.73%). Meanwhile, these three measures also exerted transregional effects on IQ decrements. More than half of the avoided IQ decrements in Hebei, Shandong and Jiangsu were associated with Hg reductions from CFPPs in other provincial regions (see details in SI Table S4). Health benefits were subject to not only Hg deposition but also the population density in each province. The amount of Hg deposited over Guangdong was comparable to that deposited over Guizhou, but the health benefits gained by the former were 5 times greater than those gained by Guizhou. The striking difference can be explained by the fact that the population in Guangdong was much larger than that in Guizhou. Furthermore, Sichuan is one of the most densely populated provinces, leading to greater health benefits in Sichuan than in less populated provinces under the same deposition reduction level. For the same reason, the health benefits gained by Inner Mongolia and Shaanxi were mismatched with their contributions to emission reductions (i.e., 2 deaths prevented locally versus 14 deaths prevented in other regions).

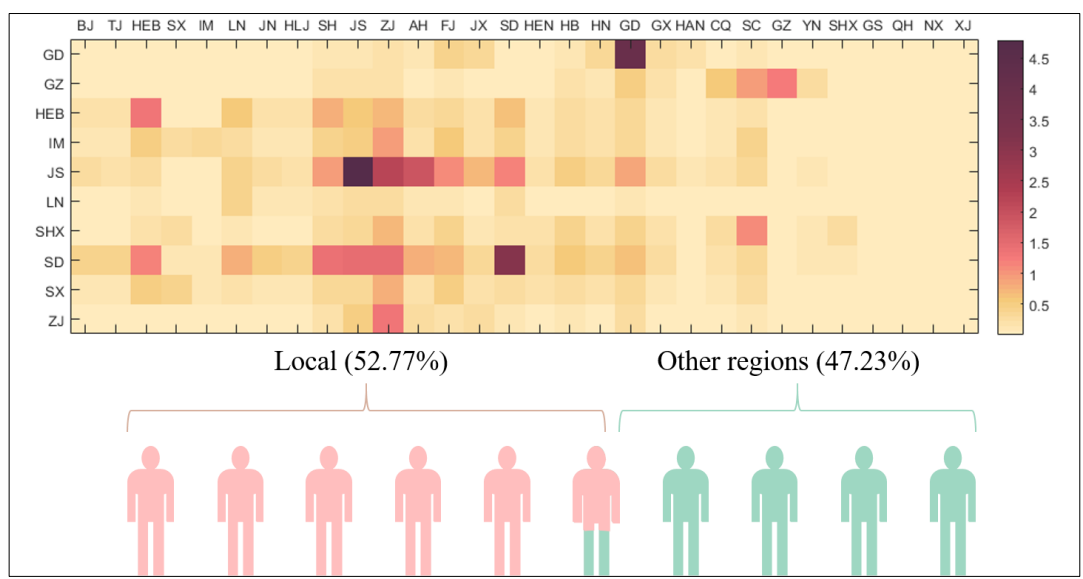

Figure 4 Source-receptor relationship of health benefits (Beijing (BJ); Tianjin (TJ); Hebei (HEB); Shanxi (SX); Inner Mongolia (IM); Liaoning (LN); Jilin (JL); Heilongjiang (HLJ); Shanghai (SH); Jiangsu (JS); Zhejiang (ZJ); Anhui (AH); Fujian (FJ); Jiangxi (JX); Shandong (SD); HEN (Henan); Hubei (HB); Hunan (HN); Guangdong (GD); Guangxi (GX); Hainan (HAN); Chongqing (CQ); Sichuan (SC); Guizhou (GZ); Yunnan (YN); Xizang (XZ); Shaanxi (SHX); Gansu (GS); Qinghai (QH); Ningxia (NX); Xinjiang (XJ).)

**Discussion**

For the first time, we performed a quantitative analysis of the Hg reduction-related health benefits gained by unit-specific CFPP retrofitting measures during the 12$^{th}$ FYP in China by combining plant-level Hg emission inventories and the China Hg Risk Source-Tracking Model. The results indicated that China has achieved significant health benefits from Hg reductions by CFPP retrofitting, with 23.51 tons Hg emissions (17.14 tons Hg$^0$, 6.35 tons Hg$^{2+}$ and 0.19 tons Hg$_p$) avoided during the study period; this value is equivalent to nearly one-fifth of the Hg released from China's CFPPs in 2010. The reductions in Hg emissions prevented 0.0021 points of per-foetus IQ decrements and 114 deaths from fatal heart attacks. The health benefits of the three different measures applied to CFPPs in China as a whole and in each provincial region were also identified. Consequently, this study can serve as an integral part of China's regular report of the progress in Hg mitigation actions, which is required by the Minamata Convention. Moreover, the present study not only provides insights for formulating unit-specific retrofitting schemes for the CFPP sectors and other important Hg emitters but also enables the administration offices to develop and adjust the Hg mitigation roadmap.

The results demonstrate that health benefits associated with the reduced Hg due to PGE improvement and the installation of APCDs mainly occurred in large CFPPs with capacities over 300 MW. On the one hand, small units, especially those with capacities under 300 MW, that failed to meet the increasingly stringent pollution emission standards were phased out; on the other hand, the costs of efficiency improvements and the installation of APCDs are beyond the affordability of small units. For example, the costs will increase by approximately 0.01 Yuan per kilowatt hour if a power plant adopts activated carbon injection for Hg reduction[20]. In addition, improving the PGE by upgrading the boiler consumes substantial investments. In view of the low capability of small units to afford technology innovations and equipment upgrades, large CFPPs should be prioritized in the expensive goals of PGE improvement and the application of efficient APCDs. Moreover, priority should be given to CFPPs owned by the five state-owned power generation groups, as they have favourable characteristics such as high monetary capital for CFPP retrofitting. According to the "Ultra-Low Emission and Energy Saving of Coal-fired Power Plant Plan", the capacity of newly built CFPPs after 2014 must be above 600 MW, and energy efficient technologies such as ultra-supercritical boilers should be introduced. Thus, large CFPPs are predicted to dominate and make high contributions to Hg reductions and their related health benefits by improving the efficiency of coal combustion and increasing the application rate of APCDs in the foreseeable future. Owing to the inherent characteristics of existing equipment, including steam pressure, temperature and combustion mode, the coal consumption rate may not be radically improved in a short time period, but the benefits of Hg emission reductions will emerge gradually over a long period[13]. Notably, phasing out recently built small units that still have a long service life and were installed with relatively adequate APCDs will waste money and resources. Therefore, it is encouraged to transform these small units into back-pressure heating units to improve their overall energy utilization efficiency.

Incentivized by the increasingly stringent environmental requirements, China now is committing to the efficient and clean utilization of coal resource, especially for power plants. Despite SUS has made remarkable contributions to reducing Hg emissions as well as health impacts during the 12th FYP period, there is still a long way to go. The total capacity of China's small units by the end of 2015 was 212.5 GW, approximately equivalent to the total capacity of CFPPs in America. Since there are still a large number of small units in China, and many of which will come to the end of their service life before 2030, SUS will maintain to be the focus of pollution mitigation strategies for the next decade. From the view of provincial regions, the extant small units concentrate in several provincial regions such as Shandong (29.45 GW), Inner Mongolia (12.57 GW), Henan (10.61 GW), Jiangsu (10.23 GW), and Shanxi (9.62 GW), which together occupy approximately half of the national total. However, if all the small unit are decommissioned radically in a short time, some regions may be confronted with large power gap. For example, small units with capacity less than 300MW take over 35% of the total CFPPs in Shandong, and there will be more than 100 million kilowatt hours shortage for power generation, if all the small units are decommissioned immediately. To solve this problem, a reasonable timetable for provincial CFPP shutdown campaign in China is recommended in consideration of the lifetime of small units in each provincial region. For the provinces with abundant renewable energy resources, such as Inner Mongolia, Gansu and Qinghai, their local renewable energy could be further explored as a substitution for coal power. For provinces such as Shandong and Jiangsu, where coal power dominated and the potential of renewable energy is comparatively smaller, replacing pollutant-intensive small units by large ones with higher Hg removal efficiency could be a feasible policy for Hg emission control. Moreover, the distributed micro-grid, energy storage system and comprehensive utilization system of power generation could be developed to prevent wind and solar curtailment.

Our analysis also reveals that mitigating Hg via retrofitting CFPPs produces significant mutual health benefits between different regions (see Figure 4). For example, owing to reductions in the Hg emissions of CFPPs in Zhejiang, the number of Hg-related deaths outside the border of the provincial territory is 1.5 times higher than that inside the provincial territory. This is because Hg mitigation in one region not only reduces local Hg deposition but also decreases Hg deposition over neighbouring regions via atmospheric transport. Thus, it is crucial to promote joint efforts to reduce Hg emissions from CFPPs. As mentioned above, the cost for retrofitting is so high that some poor regions may not be able to afford the financial burden, which will hinder transitions to cleaner CFPPs. Therefore, rich and developed regions are encouraged to provide financial and technological support to their underdeveloped counterparts, which have great Hg mitigation potential, to enhance their capabilities to retrofit CFPPs and industries. Following the successful implementation of "Clean Development Mechanism" projects, eastern regions can provide financial support to western coal bases with underdeveloped economies for technical transformations and upgrades, which will reduce trans-boundary Hg deposition and the corresponding health impacts. Furthermore, health benefits are associated not only with the total Hg reduction but also with the population density of each province. It has been verified that more densely populated areas suffer greater health impacts than less densely populated areas under the same amount of pollutant emissions[3]. For instance, the reduction in the Hg deposition of Sichuan is approximately the same as that of Gansu, but the number of avoided deaths in Sichuan is 5 times higher than that in Gansu, as the population in Sichuan is more than 2 times greater than that in Gansu. This fact implies that, when selecting sites for CFPPs or factories with large air pollutant emissions, population density should be taken into consideration. To reduce potential health impacts, the densely urbanized Beijing-Tianjin-Hebei Region, Yangtze River Delta and Pearl River Delta should prohibit the construction of new CFPPs, except for cogeneration projects, which is consistent with the "Ultra-Low Emission and Energy Saving of Coal-fired Power Plant Plan". Moreover, China's western regions have sparse populations and rich coal resources, while the opposite is true in central and eastern China, so large coal power bases should be established in the western region to expand the scale of coal power transmissions from west to east. Therefore, when formulating the national air pollutant emission reduction strategy, the resource endowments, installed capacities and power structures among different regions should be fully considered.

We also found that during the application of new APCDs (e.g., SCR), a few regions (e.g., Anhui) experienced an undesired outcome, and the amount of $Hg^{2+}$ and $Hg_P$ emissions rose with decreasing THg emissions. Increases in $Hg^{2+}$ and $Hg_P$ cause increased human health impacts, as the possibility of the intake of MeHg increases. This undesirable situation suggests that when selecting Hg control devices, policy makers should prioritize the reduction of human health impacts rather than the reduction of THg. However, the latest pollutant emission standards for CFPPs set only a total emission reduction of 0.03 mg/m$^3$, while standards for $Hg^{2+}$ and $Hg_P$ have been lacking[21]. Thus, the emission limits of $Hg^{2+}$ and $Hg_P$ for CFPPs should also be included in environmental mandates in the future. Moreover, for those CFPPs that cause increased health impacts after the use of new control devices, measures that are able to reduce emissions of all Hg species should be implemented. For instance, if ESP is replaced by ESP-FF or ESP-WESP in the combination of SCR+WFGD+ESP, not only will the overall Hg removal efficiency be increased by 25% but also the emissions of $Hg^{2+}$ and $Hg_P$ will be effectively reduced[12]. In future work, we can follow the requirements of the Minamata Convention to conduct a comprehensive evaluation and enable stakeholders to identify the best available Hg removal devices that are not only able to reduce emissions but also are economically and technically feasible for any given CFPP.

**Materials and Methods**

**Research boundary and data sources.** The framework used by this study to estimate the avoided human health impacts of reductions in Hg emissions from CFPPs is illustrated in Fig. 5. Three types of measures for reductions in Hg emissions from CFPPs, i.e., SUS, the installation of new APCDs and PGE improvement, were considered in this study. To estimate reductions in Hg emissions during the $12^{th}$ FYP, this study selected 2010 (the last year of the $11^{th}$ FYP period) as the base year. The information on decommissioned CFPPs shown in Table S9 was derived from reports issued by the National Energy Administration[22]. Notably, this study only considered Hg emission reductions resulting from newly installed APCDs and PGE improvement during 2011-2014, as the latest available information on each unit's APCDs and the PGE was for 2014. In addition, the key parameters for plant-specific Hg emission estimation, including estimates of the amount of coal burned, the coal consumption rate and APCD application, were collected from the China Electricity Council and Ministry of Ecology and Environmental Protection[23-25].

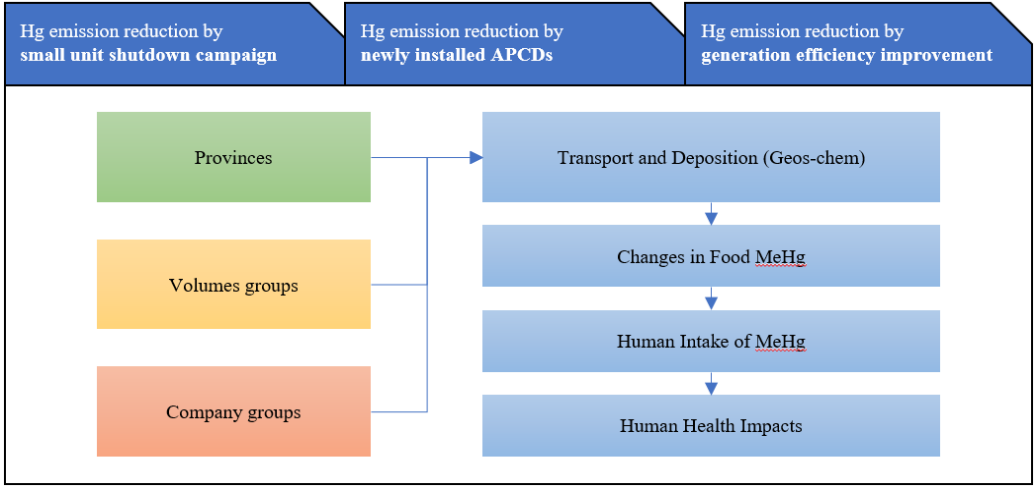

Figure 5. Framework of the Spatially Explicit Global Tracking of Hg-related Health Impacts (SEGTHI) model in this study.

***Emission reductions of small unit shutdown campaign.*** SUS generally refers to units with capacities less than 200 MW. Small units are much more pollution intensive than large units, as they lack APCDs and are installed with low-efficiency subcritical boils[13,16,26]. It was reported that the total capacity of decommissioned power plants reached 33.23 GW during the period of the 12th FYP, approximately three quarters of the capacity of Japan. Based on the accurate inventory of decommissioned CFPPs, the atmospheric Hg emissions of 234 CFPPs in 2010 was calculated as follows[16,27,28]:

$$E_{ij}^{x} = C_{ij} \times A_i \times (1 - Q_i \times \omega) \times R \times (1 - \eta_{ij}) \qquad (1)$$

where $E_{ij}$ is the Hg emissions of power plant $j$ in region $i$; $C$ is the coal consumption of plant $j$; $A_i$ is the average Hg content in consumed coal in region $i$; $Q_i$ is the percentage of washed coal in region $i$; $\omega$ is the Hg removal efficiency for coal washing; $R$ is the release ratio; and $\eta_{ij}$ represents the Hg removal efficiency of APCDs in power plant $j$ in region $i$.

***Emission reductions from newly installed APCDs.*** By the end of 2010, the CFPPs without desulfurization and denitration facilities accounted for 14% and 88% of the national total[17], respectively. In recent years, the Chinese government issued a series of policies for end-of-pipe control facilities, such as the "Emission standard of air pollutants for thermal power plants"[21] and "Ultra-Low Emission and Energy Saving of Coal-fired Power Plant Plan"[29], aiming at the large-scale application of APCDs that are able to remove Hg from flue gas[12,30]. The newly installed APCDs, including desulfurization and denitration facilities, doubled the Hg removal efficiency compared to boils only equipped with dust collectors. The calculation of the emission reductions by newly installed APCDs is expressed as follows[12,30]:

$$E_{ij}^y = C_{ij}^{t2} \times A_i \times (1 - Q_i \times \omega) \times R \times ((1 - \eta_{ij}^{t1}) - (1 - \eta_{ij}^{t2})) \quad (2)$$

where $E_{ij}^1$ is the reduction in emissions from newly installed APCDs, and $\eta_{ij}$ is the Hg removal efficiency of power plant $j$ in region $i$.

***Emission reductions by improving PGE.*** The coal consumption rate is the overall energy efficiency of power plants, implying that PGE improvements can reduce Hg emissions by reducing the coal consumption over one kilowatt hour of electric power. During the studied period, a large-scale technical reform was launched for CFPPs, and the emission reduction of 571 CFPPs was calculated with the specific parameters of this study. The amount of coal saved via the decrease in the coal consumption rate was estimated by using the following equation:

$$E_{ij}^y = P_{ij}^{t2} \times (CCR_{ij}^{t1} - CCR_{ij}^{t2}) \times A_i \times (1 - Q_i \times \omega) \times R \times (1 - \eta_{ij}^{t2}) \quad (3)$$

where $P_{ij}^{t2}$ is the electric energy production of power plant $j$ in region $i$ in $t_2$, and $CCR_{ij}$ is the coal consumption rate of power plant $j$ in region $i$.

**Atmospheric Hg biogeochemical cycle**

The methods to estimate health impacts through the Hg biogeochemical cycle in this study refer to the methods of our previous work[3]. The Hg biogeochemical cycle consists of five components: the atmospheric Hg emission inventory, an atmospheric transport model, changes in food MeHg caused by atmospheric deposition, a human intake inventory of MeHg, and human health impacts due to MeHg intake. The methods consider the Hg biogeochemical cycle to occur over mainland China and coastal seas. Hg emissions from natural sources and foreign anthropogenic sources, and imports of food products from foreign countries are also taken into consideration. In this section, we briefly describe the methods and their applications in this study, while full description of the Hg biogeochemical cycle can be found in our previous work[3].

First, Hg emission reductions in coal-fired power plants due to each of the three types of measures (i.e., SUS, the new installation of APCDs and PGE improvement) are grouped by provinces, volumes, and companies, respectively. Hg emission reductions in each group are treated as the atmospheric Hg emission inventory. We use the positions of the coal-fired power plants to spatialize the Hg emission reductions, which are used as the input of the atmospheric transport model.

Second, the GEOS-Chem chemical transport model (version 9-02; http://geos-chem.org) is used to simulate atmospheric Hg transport and deposition over China. The model includes cycling of elemental $Hg^0$, divalent $Hg^{II}$, and particulate $Hg^P$ between the atmosphere, land, and the mixed layer of the surface ocean.[31-34] A nested simulation at a horizontal resolution of 1/2°×2/3° over China was conducted in the global transport model.[35] A global simulation at a horizontal resolution of 4°×5° provided boundary conditions to the nested simulation. Both simulations were driven by GEOS-5 assimilated meteorological data from the NASA Global Modeling and Assimilation Office (GMAO). We conducted a 3-year simulation for 2008-2010 with 2008-2009 for spin-up and 2010 for analysis. The use of the same simulated time interval for each emission scenario isolated the impacts of emissions that were not affected by meteorological factors. Through evaluation against a series of observations, the nested model improved its performance over China.

Third, we estimated changes in Hg concentrations in Hg-containing food products based on the simulated changes in Hg deposition. Hg concentrations in 10 categories of food products were estimated according to previous studies[36,37]. We assumed that changes in MeHg concentrations in food products were proportional to changes in the atmospheric inputs of total Hg (THg) to environmental media, which was consistent with previous studies[38,39].

Fourth, we compiled a MeHg intake inventory for the population of China, considering 1) MeHg concentrations in food products, 2) the interprovincial trade of food products, and 3) the evaluation of the estimated daily intake (EDI). MeHg concentrations in food products of different provinces were obtained from previous studies. The multiregional input-output (MRIO) model was used to investigate the interprovincial trade of food products. The EDI of MeHg was calculated with the MeHg concentrations and the intake rate of each category of food product.

Finally, we evaluated the health impacts of Hg emission reductions in coal-fired power plants. Foetal IQ decrements and deaths from fatal heart attacks were evaluated with their dose-response relationships with dietary intake of MeHg, which were based on previous epidemiological studies[40-44]. A detailed description of the dose-response relationships and their coefficients is provided in our previous study[3].

# References

1. Giang, A. & Selin, N. E. Benefits of mercury controls for the United States. *Proceedings of the National Academy of Sciences of the United States of America* **113**, 286-291.
2. Global Mercury Assessment 2018. (United Nations Environment Programme, Switzerland, 2019).
3. Chen, L. *et al.* Trans-provincial health impacts of atmospheric mercury emissions in China. *Nat Commun* **10**, 1484, doi:10.1038/s41467-019-09080-6 (2019).
4. Roman, H. A. *et al.* Evaluation of the Cardiovascular Effects of Methylmercury Exposures: Current Evidence Supports Development of a Dose–Response Function for Regulatory Benefits Analysis. *Environmental Health Perspectives* **119**, 607-614.
5. Driscoll, C. T., Mason, R. P., Chan, H. M., Jacob, D. J. & Pirrone, N. Mercury as a Global Pollutant: Sources, Pathways, and Effects. *Environmental Science & Technology* **47**, 4967-4983.
6. Grandjean, P., Satoh, H., Murata, K. & Eto, K. Adverse Effects of Methylmercury: Environmental Health Research Implications. *Environmental Health Perspectives* **118**, 1137-1145.
7. Wilson, S., Kindbom, K., Yaramenka, K., Steenhuisen, F. & Munthe, J. *Technical Background Report for the Global Mercury Assessment 2013*. (2013).
8. Zhang, L. *et al.* Updated Emission Inventories for Speciated Atmospheric Mercury from Anthropogenic Sources in China. *Environmental Science & Technology* **49**, 3185-3194.
9. Rothenberg, S. E., Windham-Myers, L. & Creswell, J. E. Rice methylmercury exposure and mitigation: A comprehensive review. *Environmental Research* **133**, 407-423 (2014).
10. Wu, R. *et al.* Air quality and health benefits of China's emission control policies on coal-fired power plants during 2005–2020. *Environmental Research Letters* **14**, 094016, doi:10.1088/1748-9326/ab3bae (2019).
11. National Development and Reform Commission of China. Ultra-Low Emission and Energy Saving of Coal-fired Power Plant Plan (2014-2020). (Beijing, China, 2014 (in Chinese)).
12. Wu, Q., Li, G., Wang, S., Liu, K. & Hao, J. Mitigation options of atmospheric Hg emissions in China. *Environmental science & technology* **52**, 12368-12375 (2018).
13. Tong, D. *et al.* Current Emissions and Future Mitigation Pathways of Coal-Fired Power Plants in China from 2010 to 2030. *Environmental science & technology* **52**, 12905-12914 (2018).
14. Li, J., Wei, W., Zhen, W., Guo, Y. & Chen, B. How green transition of energy system impacts China's mercury emissions. *Earth's Future* (2019).
15. Liu, K. *et al.* Measure-Specific Effectiveness of Air Pollution Control on China's Atmospheric Mercury Concentration and Deposition during 2013–2017. *Environmental science & technology* **53**, 8938-8946 (2019).
16. Liu, K. *et al.* A highly resolved mercury emission inventory of Chinese coal-fired power plants. *Environmental science & technology* **52**, 2400-2408 (2018).
17. Wu, Q. *et al.* Temporal trend and spatial distribution of speciated atmospheric mercury emissions in China during 1978–2014. *Environmental science & technology* **50**, 13428-13435 (2016).
18. China Electricity Council. China Electric Power Yearbook 2011. (Beijing, China, 2011).
19. China Electricity Council. China Electric Power Yearbook 2016. (Beijing, China, 2016).

20  Jones, A. P., Hoffmann, J. W., Smith, D. N. & Murphy, J. T. DOE/NETL's phase II mercury control technology field testing program: preliminary economic analysis of activated carbon injection. *Environmental Science & Technology* **41**, 1365-1371 (2007).

21  Ministry of Ecology and Environmental of the People's Republic of China. Emission standard of air pollutants for thermal power plants. (Beijing, China, 2011).

22  National Energy Administration. The inventory of decommissioned coal-fired power plants (2011-2015). (Beijing, China, 2011-2015).

23  China Electricity Council. China Electric Power Yearbook 2011 (2010-2015). (Beijing, China, 2011-2016).

24  Ministry of Ecology and Environmental of the People's Republic of China. The inventory of desulfurization facilities of coal-fired power plants in China. (Beijing, China, 2014).

25  Ministry of Ecology and Environmental of the People's Republic of China. The inventory of denitration facilities of coal-fired power plants in China. (Beijing, China, 2014).

26  Zhao, Y. *et al.* Primary air pollutant emissions of coal-fired power plants in China: Current status and future prediction. *Atmospheric Environment* **42**, 8442-8452 (2008).

27  Zhu, C. *et al.* A high-resolution emission inventory of anthropogenic trace elements in Beijing-Tianjin-Hebei (BTH) region of China. *Atmospheric environment* **191**, 452-462 (2018).

28  Zhou, S. *et al.* Impact of a Coal-Fired Power Plant Shutdown Campaign on Heavy Metal Emissions in China. *Environ Sci Technol* **53**, 14063-14069, doi:10.1021/acs.est.9b04683 (2019).

29  National Energy Administration. Ultra-Low Emission and Energy Saving of Coal-fired Power Plant Plan. (2014).

30  Tian, H. *et al.* Control strategies of atmospheric mercury emissions from coal-fired power plants in China. *Journal of the Air & Waste Management Association* **62**, 576-586 (2012).

31  Selin, N. E. *et al.* Global 3-D land-ocean-atmosphere model for mercury: Present-day versus preindustrial cycles and anthropogenic enrichment factors for deposition. *Global Biogeochemical Cycles* **22**, GB2011, doi:10.1029/2007GB003040 (2008).

32  Holmes, C. D. *et al.* Global atmospheric model for mercury including oxidation by bromine atoms. *Atmos. Chem. Phys.* **10**, 12037-12057, doi:10.5194/acp-10-12037-2010 (2010).

33  Soerensen, A. L. *et al.* An improved global model for air-sea exchange of mercury: High concentrations over the North Atlantic. *Environmental science & technology* **44**, 8574-8580 (2010).

34  Amos, H. M. *et al.* Gas-particle partitioning of atmospheric Hg(II) and its effect on global mercury deposition. *Atmos. Chem. Phys.* **12**, 591-603, doi:10.5194/acp-12-591-2012 (2012).

35  Chen, L. *et al.* Trans-provincial health impacts of atmospheric mercury emissions in China. *Nature Communications* **10**, 1484, doi:10.1038/s41467-019-09080-6 (2019).

36  Zhang, H., Feng, X., Larssen, T., Qiu, G. & Vogt, R. D. In inland China, rice, rather than fish, is the major pathway for methylmercury exposure. *Environ. Health Perspect.* **118**, 1183-1188, doi:10.1289/ehp.1001915 (2010).

37  Liu, M. *et al.* Impacts of farmed fish consumption and food trade on methylmercury exposure in China. *Environ. Int.* **120**, 333-344, doi:https://doi.org/10.1016/j.envint.2018.08.017 (2018).

38  Meng, B. *et al.* The process of methylmercury accumulation in rice (Oryza sativa L.). *Environ. Sci. Technol.* **45**, 2711-2717, doi:10.1021/es103384v (2011).

39  Zhao, L. *et al.* Mercury methylation in rice paddies and its possible controlling factors in the Hg mining area, Guizhou province, Southwest China. *Environ. Pollut.* **215**, 1-9, doi:https://doi.org/10.1016/j.envpol.2016.05.001 (2016).

40  Giang, A. & Selin, N. E. Benefits of mercury controls for the United States. *Proc. Natl. Acad. Sci. U. S. A.* **113**, 286-291, doi:10.1073/pnas.1514395113 (2016).

41  Roman, H. A. *et al.* Evaluation of the cardiovascular effects of methylmercury exposures: current evidence supports development of a dose-response function for regulatory benefits analysis. *Environ. Health Perspect.* **119**, 607-614, doi:10.1289/ehp.1003012 (2011).

42  Rice, G. E., Hammitt, J. K. & Evans, J. S. A Probabilistic Characterization of the Health Benefits of Reducing Methyl Mercury Intake in the United States. *Environ. Sci. Technol.* **44**, 5216-5224, doi:10.1021/es903359u (2010).

43  Axelrad, D. A., Bellinger, D. C., Ryan, L. M. & Woodruff, T. J. Dose-response relationship of prenatal mercury exposure and IQ: an integrative analysis of epidemiologic data. *Environ. Health Perspect.* **115**, 609-615, doi:10.1289/ehp.9303 (2007).

44  Guallar, E. *et al.* Mercury, Fish Oils, and the Risk of Myocardial Infarction. *N. Engl. J. Med.* **347**, 1747-1754, doi:10.1056/NEJMoa020157 (2002).